\documentclass{aa}  

\usepackage{graphicx}
\usepackage{txfonts}
\usepackage{natbib}

\usepackage[colorlinks=true,
   linkcolor=blue, citecolor=blue, filecolor=blue, urlcolor=blue]{hyperref}

\defcitealias{Alarcon2026}{Paper~I}

\begin{document}

\title{A dedicated survey of fast-rotating near-Earth asteroids\\
with the Two-meter Twin Telescope}
\subtitle{II. Tumblers}

   \author{M.R. Alarcon\inst{1,2,3}\fnmsep\thanks{Corresponding author: mra@mralarcon.com}
        \and D.J. Scheeres\inst{4}
        \and J. Licandro\inst{2,3}
        \and C.J. Benson\inst{4}
        \and M. Serra-Ricart\inst{1,2,3}
        }
   \authorrunning{Alarcon et al.}
   \titlerunning{A dedicated survey of fast-rotating NEAs: tumblers} 
   
   \institute{
        Light Bridges, Observatorio Astronómico del Teide. Carretera del Observatorio del Teide, s/n, E-38570 Güímar, Tenerife, Spain \and 
        Instituto de Astrofísica de Canarias (IAC), C/ Vía Láctea, s/n, E-38205, La Laguna, Tenerife, Spain \and
        Departamento de Astrofísica, Universidad de La Laguna (ULL), E-38206 La Laguna, Tenerife, Spain \and
        University of Colorado Boulder, 3775 Discovery Drive, Boulder, CO 80303, USA
            }

   \date{Received September 7, 2026}

\abstract 
{Fast rotation dominates the metre- to decametre-size near-Earth asteroid (NEA) population, but the incidence of non-principal-axis (NPA) rotation, or tumbling, in this regime is essentially unconstrained.}
{We establish how common NPA rotation is among small, fast-rotating NEAs, how its prevalence varies with size, and what it implies for the internal structure and recent dynamical history of these bodies.}
{We analysed the homogeneous high-cadence sample obtained with the Two-meter Twin Telescope at Teide Observatory and presented in the first paper of this series. Light curves showing coherent residual structure under a single-frequency Fourier model were refitted with a bivariate Fourier series in two independent frequencies, and dynamical modes were assigned from the normalised harmonic amplitude spectrum.}
{We confirm 32 NPA rotators, 14 with both periods reliably resolved, with periods spanning 0.48--46.4 min and estimated diameters of 7--140 m. The most extreme object with both periods resolved, 2025 FB8 (shortest period $P_1 = 71$~s, estimated diameter $D \simeq 7$~m), is the fastest and smallest such tumbler characterised to date. Eight objects satisfy our harmonic quality criteria, and all eight are consistent with long-axis mode rotation. The tumbler fraction among fast rotators is 16.0\%, rising from 13.8\% at absolute magnitude $H < 26$ to 22.5\% at $H \geq 26$. Twenty-three of the 32 lie above the spin limit of a cohesionless rubble pile; none exceeds the lunar-regolith limit.}
{Tumbling is common among small fast rotators, with no evidence that it becomes rarer towards the smallest sizes. Fast tumblers require modest cohesion but no monolithic interior, occupying the same structural domain as their principal-axis counterparts.}

   \maketitle

\nolinenumbers

\section{Introduction}
\label{sec:intro}
 
The rotational state of a near-Earth asteroid (NEA) provides a direct probe of its internal structure, the torques that act on it, and the recent dynamical history that has shaped both. In the metre- to decametre-size regime, fast rotation is the rule rather than the exception: in \citet{Alarcon2026} (hereafter \citetalias{Alarcon2026}) we presented a dedicated high-cadence survey of small NEAs and identified 156 fast rotators with $P < 2.2$~h among the 249 observed targets, spanning absolute magnitudes $22 \lesssim H \lesssim 30$ and estimated diameters from a few metres to $\sim 140$~m. The fast-rotator fraction rises steeply with decreasing size, reaching more than $94\%$ of all targets with $H > 26$. Such short periods exceed the spin limit of cohesionless rubble piles \citep{Pravec2000,Holsapple2007} and require either non-negligible material strength or coherent monolithic interiors to prevent centrifugal disruption \citep{Scheeres2010,Scheeres2018}. Fast rotation in this size range is therefore not a marginal phenomenon but a population-level feature whose origin and persistence call for an explanation.
 
The dominant non-collisional process driving the long-term spin evolution of small NEAs is the Yarkovsky--O'Keefe--Radzievskii--Paddack effect (YORP): a secular torque arising from the anisotropic reflection and thermal re-emission of absorbed sunlight by an irregularly shaped body \citep{Rubincam2000}. Its characteristic timescale decreases steeply with size and heliocentric distance, so for metre- to decametre-scale NEAs, YORP is expected to act on timescales shorter than their collisional and dynamical lifetimes \citep{RozitisGreen2013}. Although equilibrium statistical models predict equal probabilities for spin-up and spin-down, observational evidence to date is dominated by accelerating cases \citep{Durech2024}, supporting a picture in which YORP routinely drives small NEAs towards and beyond the cohesionless spin barrier.
 
Superposed on this secular evolution, several short-lived processes can excite departures from principal-axis (PA) rotation and push an asteroid into a non-principal-axis (NPA), or tumbling, state. Sub-catastrophic impacts and close planetary encounters can change the spin state abruptly. Internal mass redistribution, reshaping, or episodic mass shedding near the spin limit can also trigger NPA rotation \citep[e.g.][]
{Scheeres2000, Henych2013, Pravec2014, Benson2023}. Once excited, NPA rotation is expected to be damped by internal dissipation, which removes the excess rotational energy and drives relaxation back towards the minimum-energy PA state \citep{Sharma2005,Frouard2018}. Because the damping timescale scales steeply with rotation period, this process is expected to be especially efficient at the very short periods studied here, naturally suggesting that NPA rotation should be uncommon among fast rotators.
 
Observationally, however, the picture is far from settled. Well-characterised fast-rotating tumblers remain scarce in the literature: the Asteroid Lightcurve Database \citep[LCDB;][]{Warner2009} lists only 12 tumblers among 505 fast rotators with a quality code $U\geq 2$. Dedicated campaigns have added a limited number of cases, including 10 tentative NPA candidates in the Mission Accessible Near-Earth Objects Survey \citep[MANOS;][]{Thirouin2018} and 9 more (5 with measured periods) reported by \citet{Sanchez2024}. These numbers are too small and heterogeneous in instrumentation, cadence, and selection criteria to place meaningful constraints on the true prevalence of tumbling in the fast-rotation regime. A particularly instructive individual case is the $\sim$10~m NEA 2012~TC$_4$ \citep{Lee2021}, for which time-series photometry obtained during its 2012 and 2017 close approaches revealed an evolving tumbling state. The change in rotation and precession periods between the two apparitions could not be explained by the Earth flyby, internal dissipation, or interplanetary impacts alone, making YORP acting on a tumbling body the most plausible driver. Theoretically, radiative torques on tumbling bodies are known to differ from the PA case and, when combined with internal dissipation, can sustain long-lived excited configurations for some parameter ranges \citep{Vokrouhlicky2007,Breiter2015,Benson2022}. The interplay between YORP-driven evolution and NPA dynamics in the small-NEA regime therefore remains insufficiently constrained, both observationally and theoretically.
 
In this work we exploit the homogeneous fast-rotator sample of \citetalias{Alarcon2026} to characterise NPA rotation in the fast-rotation regime. The dynamical framework of NPA rotation and its photometric signature are reviewed in Sect.~\ref{sec:tumbling}, the sample and the bivariate period search are described in Sect.~\ref{sec:methods}, and the resulting tumbler subsample is presented in Sect.~\ref{sec:results}. The implications of these results for the coexistence of fast rotation and sustained NPA motion are then discussed in Sect.~\ref{sec:discussion}, and the main conclusions are summarised in Sect.~\ref{sec:conclusions}.

\section{Tumbling: Dynamical framework and photometric signature}
\label{sec:tumbling}

The rotational state of an asteroid is governed by $\boldsymbol{L} = \hat{I} \boldsymbol{\omega}$, where $\boldsymbol{L}$ is the angular momentum vector, $\hat{I}$ the inertia tensor and $\boldsymbol{\omega}$ the angular velocity. For a rigid body with principal moments of inertia $I_1 \leq I_2 \leq I_3$, pure rotation about any of the three principal axes--with $\boldsymbol{\omega}$ parallel to $\boldsymbol{L}$--is referred to as PA rotation. At fixed $|\boldsymbol{L}|$, the minimum-energy state corresponds to PA rotation about the axis of maximum inertia $I_3$. Rotation about $I_1$ is also stable for a perfectly rigid body, but in the presence of internal energy dissipation $I_3$ is the only asymptotically stable PA configuration. Any departure from this state results in NPA rotation, in which $\boldsymbol{\omega}$ is no longer parallel to $\boldsymbol{L}$ and the body undergoes a complex, quasi-periodic motion commonly referred to as tumbling \citep{Harris1994}. This motion is naturally described by two fundamental timescales: a rotation period $P_\psi$ associated with the spin about a body-fixed axis and a precession period $P_\phi$ associated with the wobble of that axis around the (inertially fixed) angular-momentum vector \citep{Kaasalainen2001}.

\begin{figure}
  \centering
  \includegraphics[width=\columnwidth]{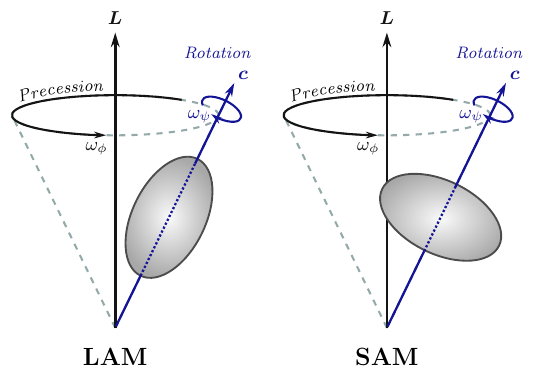}
  \caption{Schematic illustration of the two NPA rotation modes. In both cases the body axis $\boldsymbol{c}$ precesses around the fixed angular momentum vector $\boldsymbol{L}$ with angular velocity $\omega_\phi$, while the body rotates about $\boldsymbol{c}$ with angular velocity $\omega_\psi$. In LAM (\textit{left}), $\boldsymbol{c}$ is the longest principal axis of the body; in SAM (\textit{right}), $\boldsymbol{c}$ is the shortest principal axis. The figure shows that the same two-component motion---precession and rotation---gives rise to qualitatively different orientations of the body with respect to $\boldsymbol{L}$ in each mode.}
  \label{fig:lam_sam}
\end{figure}

Depending on which body axis is closer to $\boldsymbol{L}$, NPA rotation falls into one of two dynamical classes that are shown in Fig.~\ref{fig:lam_sam}. In long-axis mode (LAM), the body rotates predominantly about its longest principal axis (smallest moment of inertia, $I_1$), which precesses around $\boldsymbol{L}$. In this regime, $P_\psi/P_\phi$ can take any positive value, with no further dynamical restriction. In short-axis mode (SAM), the body rotates predominantly about its shortest principal axis (largest moment of inertia, $I_3$), again precessing around $\boldsymbol{L}$. In this case the geometry of the inertia ellipsoid imposes $P_\psi/P_\phi > 2$. A full derivation of these period ratios from the Euler equations and the constraints on the inertia ellipsoid is given in \citet{Kaasalainen2001}.

The photometric signature of tumbling reflects this bi-periodic motion. Under the assumption of constant observing geometry, which is appropriate over the relatively short observing windows considered here, the brightness of an NPA rotator can be expanded as a bivariate Fourier series in the two fundamental frequencies $f_\psi = 1/P_\psi$ and $f_\phi = 1/P_\phi$ \citep{Pravec2005}:
\begin{align}
F(t) = C_0
  &+ \sum_{j=1}^{m} \left[ C_{j0}\cos\left(2\pi j f_\psi t\right)
    + S_{j0}\sin\left(2\pi j f_\psi t\right) \right]
  \nonumber \\
  &+ \sum_{k=1}^{m}\sum_{j=-m}^{m}
    \left[ C_{jk}\cos\left(2\pi(j f_\psi + k f_\phi) t\right) \right.
  \nonumber \\
  & \qquad\qquad
    \left. + S_{jk}\sin\left(2\pi(j f_\psi + k f_\phi) t\right)
    \right],
\label{eq:bivariate}
\end{align}
where $C_0$ is the mean flux, $C_{jk}$ and $S_{jk}$ are the Fourier coefficients of the harmonic at combined frequency $jf_\psi + kf_\phi$ and $m$ is the truncation order. The set of normalised harmonic amplitudes
\begin{equation}
A^{\mathrm{norm}}_{jk} = \frac{\sqrt{C_{jk}^2 + S_{jk}^2}}{C_0}
\label{eq:anorm}
\end{equation}
characterises the relative contribution of each frequency combination to the total light-curve variation and constitutes the primary observable from which the NPA state is diagnosed.

The amplitude spectrum $\{A^{\mathrm{norm}}_{jk}\}$ encodes information about the dynamical mode of the rotation. LAM states are characterised by prominent signal at $2f_\phi$ and $2(f_\phi + f_\psi)$, whereas SAM states display prominent signal at $2f_\phi$ and $2(f_\phi - f_\psi)$ \citep{Kaasalainen2001,Pravec2005}. The dominant peak is not necessarily $2f_\phi$ in either case, as the relative strengths depend on the body geometry and viewing aspect. The unambiguous assignment of the two fitted frequencies to $f_\phi$ and $f_\psi$ requires either physical modelling of the spin state and shape \citep{Kaasalainen2001} or photometric coverage at more than one viewing geometry \citep{Pravec2005}, neither of which is available for our single-session survey. Throughout this paper we therefore label the two fitted periods $P_1$ and $P_2$ ($P_1 \leq P_2$ by construction) and use the structure of the harmonic spectrum only to classify the dynamical mode (LAM or SAM) for the subset of objects where the spectral quality criteria described in Sect.~\ref{sec:methods} are met.

\section{Methods}
\label{sec:methods}

\subsection{Observations}
\label{sec:sample}

The observational sample was obtained as part of the high-cadence survey of small NEAs presented in \citetalias{Alarcon2026}, to which we refer the reader for a complete description of the survey strategy, instrumental setup and data-reduction pipeline. The survey was carried out with the Two-meter Twin Telescope (TTT\footnote{\url{https://ttt.iac.es/}}) at Teide Observatory, Canary Islands. The facility comprises two 0.80~m (TTT1/2) and two 2.0~m (TTT3/4) Ritchey--Chr\'etien telescopes, all equipped with the FERVOR sCMOS imager \citep{Alarcon2023} and operated through a broadband UV/IR Cut filter (400--700~nm). Targets were observed during single-night runs of typically 40--120~min, with exposure times adjusted to the predicted brightness and apparent motion of each object so as to optimise the recovery of fast rotational signals.

In \citetalias{Alarcon2026} we observed 249 small NEAs and derived reliable rotation periods for 161 of them, of which 156 ($P<2.2$~h) qualify as fast rotators. The fast-rotator sample spans $22 \lesssim H \lesssim 30$, corresponding to estimated diameters from $\sim 140$~m down to a few metres for an assumed albedo $p_V = 0.15$. Among these 156 fast rotators, 35 showed coherent residual structure inconsistent with a single-period rotational model and were retained as NPA candidates for the analysis presented here.

For each fast rotator, a single-frequency Fourier model of order four was fitted to the calibrated, geometry-corrected light curve. Candidates for NPA rotation were then identified by visual inspection of the residuals and of the data folded at the best single-period solution: objects whose folded curve showed coherent, phase-correlated structure inconsistent with random scatter--typically as cycle-to-cycle variations in amplitude or shape that no single rotational waveform could reproduce--were retained and subjected to the bivariate analysis described below, which provides the formal statistical assessment used to confirm or reject each candidate.

\subsection{Bivariate period search}
\label{sec:bivariate}

For each NPA candidate we fitted the bivariate Fourier model of Eq.~\ref{eq:bivariate} to the light curve. The initial reference frequency, $f_{\rm ref}=1/P_{\rm I}$, was obtained from the period determined in \citetalias{Alarcon2026}. In that analysis, the highest-power peak of a fourth-order Lomb--Scargle periodogram provided an initial estimate, subsequently refined by phase dispersion minimisation. Candidate periods were also assessed by visual inspection of the phase-folded light curves. Assuming predominantly shape-driven variability, we adopted solutions with two maxima and two minima per rotation rather than single-peaked solutions. We then searched for $f_2$ over a dense grid spanning from one cycle over the observational baseline to the Nyquist frequency of the median sampling cadence, with $f_1$ held fixed at its refined value. At each trial $f_2$ the full bivariate model was fitted by weighted least squares. Narrow windows around the harmonics of $f_1$ were excluded from the search to avoid spurious matches with residual single-period power. The resulting global $\chi^2$ minimum was then refined jointly in both frequencies. We adopted the direct evaluation of the bivariate $\chi^2$ over a grid of $f_2$ values rather than the pre-whitening approach, in which a single-frequency model is subtracted from the data before searching for $f_2$ in the residuals. This step can absorb part of the signal carried by the cross-terms $jf_1 \pm kf_2$ into the subtracted model, biasing the residual periodogram against the correct $f_2$ \citep{Henych2013}. The grid search instead fits the complete bivariate model at each trial $f_2$ so that all cross-terms within the adopted truncation order are accounted for simultaneously.

The truncation order $m$ of the bivariate model was selected for each object using the Bayesian information criterion \citep[BIC;][]{Schwarz1978}, evaluating $m \in \{2, 3\}$. The BIC penalises model complexity more strongly than the $F$-test, favouring the simpler model unless the improvement in $\chi^2$ is large enough to justify the additional parameters. This makes it well suited to guard against overfitting in light curves where the photometric noise is comparable to the NPA amplitude. In practice, the majority of objects were best described by $m = 3$.

The improvement of the bivariate fit over the single-frequency model was assessed with an $F$-test comparing the reduction in $\chi^2$ against the increase in the number of free parameters. The bivariate analysis was applied to all candidates flagged by the visual inspection step. Solutions for which the improvement over the single-frequency model was significant at the $p < 0.01$ level were retained as confirmed NPA rotators; the remaining candidates were reclassified as PA rotators. Following the quality classification scheme of \citet{Pravec2005}, each confirmed tumbler was assigned a principal-axis rotation (PAR) quality code based on the reliability of the bivariate solution: PAR~$= -3$ when both periods are reliably resolved (baseline covering more than five full cycles of the longer period and $\chi^2_{1f}/\chi^2_{\rm biv} \geq 1.5$), and PAR~$= -2$ when NPA rotation is detected but the second period is not fully resolved. The final confirmed sample, along with the PAR assignments, is presented in Sect.~\ref{sec:results_sample}.

A small subset of targets was observed across multiple nights, resulting in light curves with internal gaps of several hours. Bivariate Fourier models are especially sensitive to such gaps because high-order combination harmonics whose period is shorter than the gap duration can develop arbitrarily large coefficients that diverge between segments while still fitting the data within each of them. For these objects we restricted the bivariate fit to the longest contiguous segment and evaluated the fitted model across the gap to detect any divergent behaviour. The periods recovered from the best segment were retained as the reported values and the PAR code reflects the quality of that segment alone.

Formal uncertainties on both frequencies were estimated analytically by propagating the photometric uncertainties through the linearised bivariate model via the Fisher information matrix, from which frequency uncertainties were extracted as the square roots of the diagonal elements of its inverse. Period uncertainties follow from $\sigma_P = \sigma_f / f^2$. As a conservative safeguard, we imposed a minimum uncertainty of $1/(2T)$ on each frequency, where $T$ is the duration of the fitted segment. This floor corresponds to the classical frequency resolution of a dataset of duration $T$ and was applied in cases where the Fisher bound would otherwise return unrealistically small values for very densely sampled or high signal-to-noise light curves. 

\subsection{Tentative mode assignment}
\label{sec:lamsam}

As discussed in Sect.~\ref{sec:tumbling}, a complete identification of $P_\phi$ and $P_\psi$ is beyond the reach of our single-epoch photometry. We nevertheless used the harmonic structure of the bivariate fit as a tentative diagnostic of the dynamical mode, applying the following operational procedure.

For each confirmed NPA rotator we computed the normalised amplitude spectrum $\{A^{\mathrm{norm}}_{jk}\}$ following Eq.~\ref{eq:anorm}. In the framework of \citet{Kaasalainen2001}, LAM states are characterised by prominent signal at $2f_\phi$ and $2(f_\phi + f_\psi)$, while SAM states display prominent signal at $2f_\phi$ and $2(f_\phi - f_\psi)$. The relative strengths of these features depend on the body geometry and the viewing aspect, so a single-epoch spectrum provides only a partial view of the full harmonic content. With this caveat, we used the observed spectrum as a tentative diagnostic: the frequency whose second harmonic carries the larger normalised amplitude ($A^{\mathrm{norm}}_{2,0}$ or $A^{\mathrm{norm}}_{0,2}$) was associated with $f_\phi$ as an operational approximation, acknowledging that this identification can fail when a combination harmonic such as $2(f_\phi + f_\psi)$ dominates the spectrum \citep[as observed for 2002~TD$_{60}$;][]{Pravec2005}. The dynamical mode was then inferred from the relative strengths of the combination harmonics, applied in the following order. An object was
classified as SAM if it satisfies the dynamical constraint $P_\psi/P_\phi > 2$ \citep{Kaasalainen2001} and $A^{\mathrm{norm}}_{2,-2} > A^{\mathrm{norm}}_{2,2}$; it was classified as LAM if $A^{\mathrm{norm}}_{2,2} > A^{\mathrm{norm}}_{2,-2}$ and the SAM condition is not met. Objects for which neither condition is clearly satisfied were left unclassified. Given these limitations, we restricted the mode classification to PAR~$= -3$ objects and required in
addition that one frequency has a clearly stronger second harmonic than the other so that the mode assignment rests on a positive spectral diagnostic rather than on the dynamical period-ratio constraint alone.

\section{Results and analysis}
\label{sec:results}

\begin{table*}[t]
\centering
\caption{Bivariate periods and observational parameters of the TTT NPA sample.\label{tab:tumblers}}
\resizebox{\textwidth}{!}{
\begin{tabular}{l c c c c c l c c c c c c}
\hline\hline
Object & $P_1$ (min) & $P_2$ (min) & $\chi^2_{1f}/\chi^2_{\rm biv}$ & PAR & Mode & Dominant terms & $H$ & $\Delta$ (au) & $\alpha$ (deg) & $T_{\rm exp}$ (s) & Night & Tel.\\
\hline
2023 BM4 & $4.380 \pm 0.079$ & $7.30 \pm 0.22$ & $2.45$ & $-3$ &   & $3(f_{1}-f_{2})$, $2f_{2}$, $2(f_{1}-f_{2})$ & $23.7$ & $0.0448$ & $30.2$ & $10.0$ & 2026-05-20 & TTT1 \\
2023 SQ & $26.9 \pm 1.8$ & $34.3 \pm 3.0$ & $5.57$ & $-3$ &   & $2f_{1}$, $2(f_{1}-f_{2})$, $f_{1}-f_{2}$ & $25.0$ & $0.0144$ & $46.1$ & $3.5$ & 2023-09-21 & TTT2 \\
2024 OY2 & $12.14 \pm 0.32$ & $19.14 \pm 0.81$ & $1.63$ & $-3$ & LAM & $2f_{2}$, $3f_{1}$, $2f_{1}$ & $25.1$ & $0.0230$ & $53.9$ & $17.0$ & 2024-08-03 & TTT2 \\
2025 CO3 & $8.74 \pm 0.32$ & $14.43 \pm 0.86$ & $3.52$ & $-3$ &   & $2f_{2}$, $2f_{1}$, $3f_{1}$ & $23.0$ & $0.0548$ & $34.7$ & $6.0$ & 2025-08-21 & TTT1 \\
2025 FB8 & $1.183 \pm 0.017$ & $2.151 \pm 0.055$ & $8.78$ & $-3$ & LAM & $2f_{1}$, $f_{1}+f_{2}$, $2(f_{1}+f_{2})$ & $28.5$ & $0.0041$ & $42.3$ & $4.0$ & 2025-03-27 & TTT3 \\
2025 KD & $14.35 \pm 0.85$ & $23.5 \pm 2.3$ & $2.66$ & $-3$ & LAM & $f_{1}$, $2f_{1}$, $f_{1}+2f_{2}$ & $25.8$ & $0.0173$ & $37.5$ & $2.1$ & 2025-05-19 & TTT3 \\
2025 MB1 & $7.32 \pm 0.42$ & $8.07 \pm 0.51$ & $2.64$ & $-3$ &   & $f_{1}+f_{2}$, $2f_{2}$, $f_{1}+2f_{2}$ & $24.6$ & $0.0323$ & $36.3$ & $2.9$ & 2025-06-21 & TTT3 \\
2025 MR89 & $1.447 \pm 0.023$ & $5.89 \pm 0.38$ & $2.38$ & $-3$ &   & $2f_{1}$, $f_{1}-f_{2}$, $3f_{2}$ & $26.3$ & $0.0189$ & $24.9$ & $2.1$ & 2025-06-29 & TTT3 \\
2025 MZ1 & $3.103 \pm 0.076$ & $6.93 \pm 0.38$ & $4.22$ & $-3$ & LAM & $2f_{1}$, $2(f_{1}+f_{2})$, $f_{1}-f_{2}$ & $26.8$ & $0.0138$ & $36.6$ & $5.1$ & 2025-06-24 & TTT3 \\
2025 QE16 & $3.06 \pm 0.11$ & $7.63 \pm 0.69$ & $3.86$ & $-3$ & LAM & $2f_{2}$, $2(f_{1}+f_{2})$, $f_{2}$ & $25.3$ & $0.0192$ & $36.6$ & $7.2$ & 2025-09-01 & TTT3 \\
2025 SA1 & $6.51 \pm 0.49$ & $7.28 \pm 0.61$ & $3.45$ & $-3$ & LAM & $f_{2}$, $f_{1}+f_{2}$, $f_{1}$ & $26.2$ & $0.0161$ & $34.6$ & $2.3$ & 2025-09-26 & TTT3 \\
2025 SJ15 & $6.41 \pm 0.24$ & $10.76 \pm 0.69$ & $1.96$ & $-3$ & LAM & $2f_{2}$, $2f_{1}$, $f_{2}$ & $24.7$ & $0.0641$ & $15.2$ & $8.5$ & 2025-09-30 & TTT3 \\
2025 UV5 & $2.278 \pm 0.056$ & $3.45 \pm 0.13$ & $5.44$ & $-3$ &   & $3f_{2}$, $2f_{1}-f_{2}$, $2f_{1}$ & $26.4$ & $0.0111$ & $57.2$ & $1.8$ & 2025-10-29 & TTT3 \\
2026 EL2 & $17.1 \pm 1.4$ & $20.4 \pm 2.0$ & $7.07$ & $-3$ & LAM & $f_{1}+f_{2}$, $-2f_{1}+3f_{2}$, $2f_{2}$ & $26.2$ & $0.0150$ & $30.7$ & $6.2$ & 2026-03-15 & TTT3 \\
[3pt]
2024 MK & $29.5 \pm 2.3$ & \textit{$46.3 \pm 5.7$} & $11.26$ & $-2$ &   & $2f_{1}-f_{2}$, $f_{2}$, $f_{1}-f_{2}$ & $22.0$ & $0.0298$ & $101.4$ & $71.7$ & 2024-07-03 & TTT2 \\
2024 SF2 & $5.89 \pm 0.15$ & \textit{$19.3 \pm 1.7$} & $1.16$ & $-2$ &   & $2f_{1}$, $f_{2}$, $f_{1}$ & $24.4$ & $0.0356$ & $43.8$ & $10.0$ & 2024-09-28 & TST \\
2025 EK4 & $5.66 \pm 0.27$ & \textit{$31.8 \pm 8.4$} & $6.74$ & $-2$ &   & $2f_{1}$, $f_{1}-f_{2}$, $f_{1}+f_{2}$ & $24.0$ & $0.0474$ & $53.6$ & $10.0$ & 2025-03-17 & TTT3 \\
2025 QB21 & $12.57 \pm 0.52$ & \textit{$33.2 \pm 3.6$} & $4.73$ & $-2$ &   & $f_{1}-2f_{2}$, $2(f_{1}-f_{2})$, $-f_{1}+3f_{2}$ & $26.4$ & $0.0108$ & $34.0$ & $5.0$ & 2025-09-02 & TTT1 \\
2025 QD8 & $6.31 \pm 0.27$ & \textit{$18.5 \pm 2.3$} & $2.00$ & $-2$ &   & $2f_{2}$, $f_{2}$, $f_{1}-f_{2}$ & $26.1$ & $0.0195$ & $19.7$ & $15.0$ & 2025-08-31 & TTT3 \\
2025 QL7 & $19.9 \pm 1.6$ & \textit{$38.5 \pm 6.1$} & $10.02$ & $-2$ &   & $f_{2}$, $f_{1}-f_{2}$, $2f_{1}-f_{2}$ & $24.1$ & $0.0913$ & $8.8$ & $19.4$ & 2025-08-28 & TTT3 \\
2025 QM23 & $2.810 \pm 0.089$ & \textit{$18.5 \pm 3.9$} & $2.20$ & $-2$ &   & $f_{1}$, $2f_{1}$, $2f_{1}-3f_{2}$ & $26.8$ & $0.0215$ & $20.4$ & $6.4$ & 2025-09-03 & TTT3 \\
2025 QR12 & $22.5 \pm 1.4$ & \textit{$40.1 \pm 4.3$} & $2.40$ & $-2$ &   & $2f_{1}-f_{2}$, $2(f_{1}-f_{2})$, $2f_{1}-3f_{2}$ & $23.8$ & $0.0352$ & $42.7$ & $5.0$ & 2025-09-01 & TTT1 \\
2025 SE29 & $16.01 \pm 0.96$ & \textit{$46.4 \pm 8.1$} & $2.08$ & $-2$ &   & $f_{1}$, $2f_{1}-3f_{2}$, $3f_{2}$ & $22.7$ & $0.0378$ & $12.3$ & $4.0$ & 2025-10-06 & TTT1 \\
2025 UE1 & $6.43 \pm 0.45$ & \textit{$22.1 \pm 5.3$} & $4.99$ & $-2$ &   & $f_{1}-f_{2}$, $2f_{2}$, $3f_{2}$ & $26.6$ & $0.0085$ & $28.8$ & $10.0$ & 2025-10-21 & TTT3 \\
2025 UU & $10.3 \pm 1.2$ & \textit{$14.6 \pm 2.4$} & $1.50$ & $-2$ &   & $-f_{1}+2f_{2}$, $2(f_{1}-f_{2})$, $f_{1}-f_{2}$ & $25.7$ & $0.0292$ & $29.4$ & $4.7$ & 2025-10-21 & TTT3 \\
2025 WC5 & $6.29 \pm 0.45$ & \textit{$16.9 \pm 3.3$} & $4.18$ & $-2$ &   & $2f_{2}$, $-f_{1}+3f_{2}$, $f_{1}-2f_{2}$ & $24.9$ & $0.0400$ & $21.0$ & $8.1$ & 2025-11-29 & TTT3 \\
2025 WT3 & $6.90 \pm 0.54$ & \textit{$13.4 \pm 2.0$} & $2.41$ & $-2$ &   & $2f_{1}-f_{2}$, $f_{1}+f_{2}$, $f_{2}$ & $25.6$ & $0.0516$ & $3.0$ & $14.4$ & 2025-12-10 & TTT3 \\
2025 YM9 & $0.478 \pm 0.023$ & \textit{$3.7 \pm 1.4$} & $2.84$ & $-2$ &   & $2f_{1}$, $f_{1}$, $f_{1}+2f_{2}$ & $28.7$ & $0.0062$ & $20.9$ & $2.0$ & 2025-12-31 & TTT3 \\
2026 AE & $1.921 \pm 0.044$ & \textit{$14.3 \pm 2.4$} & $53.11$ & $-2$ &   & $2f_{2}$, $3(f_{1}-f_{2})$, $3f_{1}+f_{2}$ & $25.2$ & $0.0484$ & $31.0$ & $40.0$ & 2026-01-10 & TTT3 \\
2026 AS2 & $12.87 \pm 0.78$ & \textit{$40.0 \pm 7.5$} & $1.86$ & $-2$ &   & $f_{1}$, $f_{1}-3f_{2}$, $2f_{1}$ & $24.2$ & $0.0299$ & $11.0$ & $4.0$ & 2026-01-25 & TTT1 \\
2026 BH2 & $4.81 \pm 0.26$ & \textit{$12.2 \pm 1.7$} & $4.12$ & $-2$ &   & $2f_{1}$, $2(f_{1}-f_{2})$, $2f_{2}$ & $26.6$ & $0.0269$ & $38.0$ & $21.4$ & 2026-01-22 & TTT3 \\
2026 JM1 & $0.878 \pm 0.011$ & \textit{$11.0 \pm 1.7$} & $1.91$ & $-2$ &   & $2f_{1}$, $f_{1}$, $2f_{1}+f_{2}$ & $27.6$ & $0.0104$ & $25.1$ & $5.0$ & 2026-05-10 & TTT4 \\
\hline
\end{tabular}
}
\tablefoot{For each object we report the two independent periods $P_1 \leq P_2$ recovered by the bivariate Fourier analysis, with formal $1\sigma$ uncertainties, together with the improvement of the bivariate fit over the single-period model ($\chi^2_{1f}/\chi^2_{\rm biv}$) and a quality code PAR$^a$ indicating whether both periods are reliably resolved. For PAR~$=-2$ objects $P_2$ is tentative. Where the harmonic structure of the bivariate solution meets strict quality criteria, a tentative spin-state classification (Mode$^b$, LAM or SAM) is also given. Dominant terms are listed in decreasing order of fitted harmonic amplitude. For PAR~$=-2$ objects, terms involving the tentative second frequency should be interpreted with caution.
\tablefoottext{a}{PAR~$=-3$: both periods reliably resolved (baseline $>5$ cycles of $P_2$, $\chi^2_{1f}/\chi^2_{\rm biv} \geq 1.5$). PAR~$=-2$: NPA rotation detected but $P_2$ not fully resolved.}\tablefoottext{b}{LAM (long-axis mode) or SAM (short-axis mode), assigned from the harmonic structure of the bivariate solution; blank where quality criteria are not met.}$H$: absolute magnitude. $\Delta$: geocentric distance (au). $\alpha$: solar phase angle (deg). Both evaluated at the median epoch of the fitted segment. $T_{\rm exp}$: individual frame exposure time (s).}
\end{table*}

\subsection{The tumbler sample}
\label{sec:results_sample}
 
The bivariate analysis was applied to the 35 NPA candidates initially identified in \citetalias{Alarcon2026}, supplemented by seven additional objects observed with the same survey strategy after the submission of that
paper. Of these 42 candidates, 32 are confirmed here as NPA rotators and 10 are reclassified as PA rotators: 14 confirmed objects have PAR~$= -3$ (both periods reliably resolved) and 18 have PAR~$= -2$ (NPA rotation detected but second period not fully resolved). The two recovered periods $P_1$ and $P_2$ for all 32 are listed in Table~\ref{tab:tumblers}, together with the formal period uncertainties, the $\chi^2$ improvement, the PAR code and the tentative mode classification where available. $P_1$ ranges from 0.48~min (2025~YM9) to 29.5~min (2024~MK) and $P_2$ from 2.15~min (2025~FB8) to 46.4~min (2025~SE29). The BIC selected $m=3$ for 23 objects and $m=2$ for the remaining 9. The median $\chi^2_{1f}/\chi^2_{\rm biv} \simeq 3.1$, reaching $\sim 53$ for 2026~AE. The 18 tumblers classified as PAR~$= -2$ are confirmed NPA rotators. However, their second periods remain poorly constrained, due to either insufficient rotational coverage within our single-night baselines or high photometric noise. The 32 tumblers presented here more than double the number of fast-rotating NPA asteroids for which two periods have been recovered in the literature to date, making this the largest homogeneous sample of fast-rotating tumblers characterised so far. We refer to these objects as fast tumblers, i.e. confirmed NPA rotators whose shorter period $P_1$ is below 2.2~h.

\begin{figure*}
    \centering
    \includegraphics[width=\textwidth]{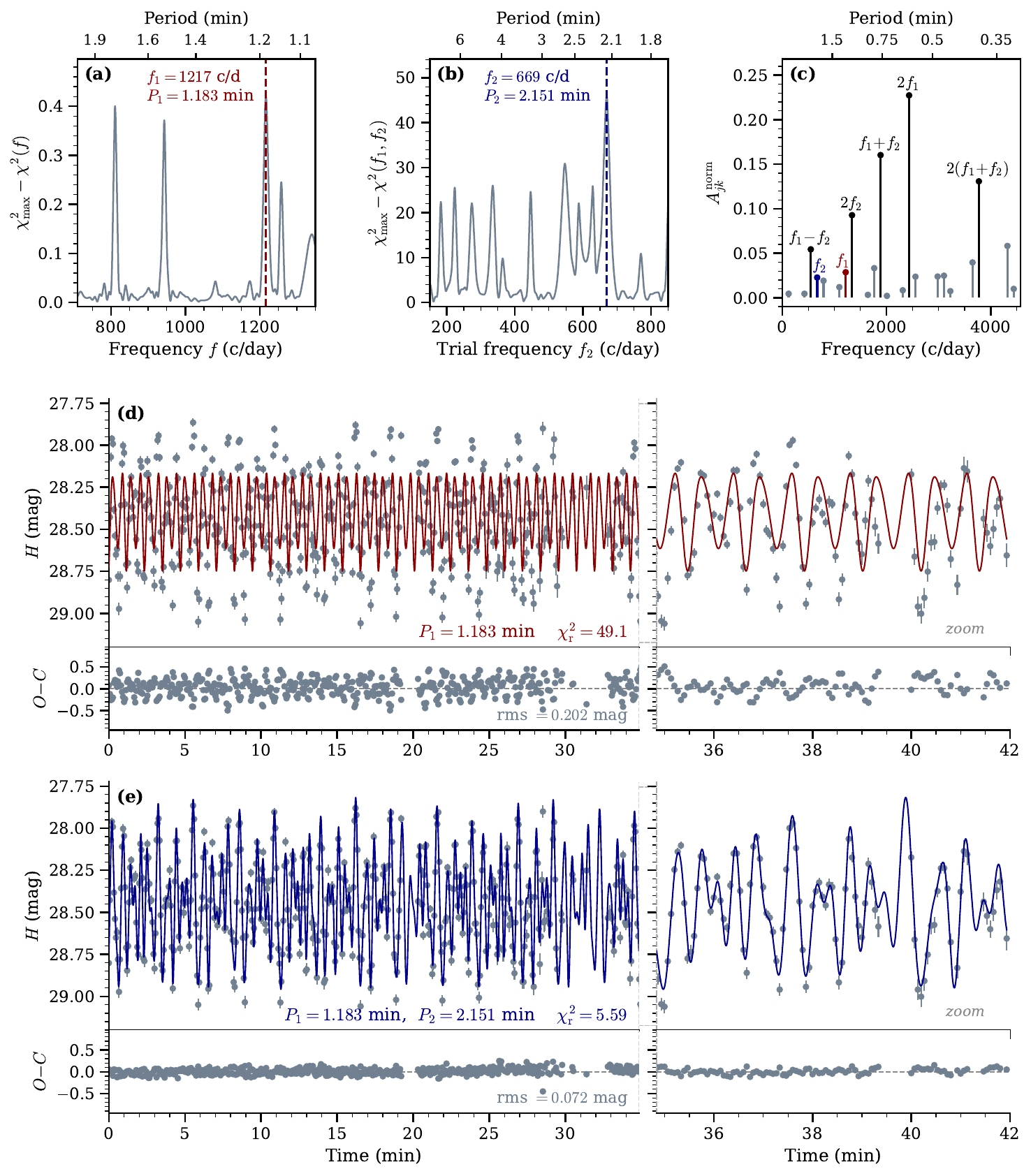}
    \caption{Bivariate analysis for 2025~FB8 ($P_1 = 1.183 \pm 0.017$~min, $P_2 = 2.151 \pm 0.055$~min, PAR~$= -3$).
\textit{(a)} Single-frequency $\chi^2$ periodogram. The more prominent peak at $f_1 = 1217$~c/d is marked.
    \textit{(b)} Bivariate $\chi^2$ grid as a function of the trial second frequency, $f_2$. The global minimum at $f_2 = 669$~c/d is indicated by the dashed blue line.
    \textit{(c)} Normalised amplitude spectrum, $A^{\rm norm}_{jk}$. Harmonics are identified by indices $(j,k)$. The dominant features $2f_1$ and $2(f_1+f_2)$ identify the object as a tentative LAM rotator.
    \textit{(d)} Reduced light curve with the best single-period model (red) and its residuals. The slow modulation at $P_2$ is visible in the $O{-}C$ panel and in the zoom inset.
    \textit{(e)} Same data but with the best bivariate model (blue) and its residuals. The modulation is fully removed.
    The units c/day and c/d denote cycles per day. Panels~(a) and~(b) show enlarged portions of the base-frequency searches, whereas panel~(c) shows the amplitudes of individual harmonic and combination terms calculated from the fitted bivariate coefficients; the displayed limits in panels~(a) and~(b) are not the full search limits.}
    \label{fig:2025FB8}
\end{figure*}
 
The full bivariate analysis is illustrated in Fig.~\ref{fig:2025FB8} for 2025~FB8 ($P_1 = 1.183 \pm 0.017$~min, $P_2 = 2.151 \pm 0.055$~min, baseline spanning 19.5 cycles of $P_2$). Panel~(a) confirms the existence of a primary period $P_1$ in the single-frequency periodogram. The key step is panel~(b): the bivariate $\chi^2$ grid identifies $f_2$ as a clear global minimum among multiple local minima corresponding to beat frequencies $|jf_1 \pm kf_2|$. Panel~(c) displays the harmonic content of the fitted bivariate model. The normalised amplitude spectrum is dominated by $2f_1$, $(f_1+f_2)$ and $2(f_1+f_2)$, the harmonics expected for LAM rotation under the framework of \citet{Kaasalainen2001}, with the remaining power distributed in lower-order combinations as predicted by the theory. The validation of the solution comes in panels~(d) and~(e): the single-period model leaves a coherent residual on the timescale of $P_2$ with $\chi^2_r = 49.1$, and adding $f_2$ to the model collapses this residual to $\chi^2_r = 5.59$. The bivariate solution is therefore not a mathematical artefact of adding parameters, but the explicit recovery of a second physical frequency present in the data.
 
While our sample includes one object with a shorter $P_1$ and a smaller size (2025~YM9, with $P_1 = 29$~s and $H = 28.7$), 2025~FB8 stands as the most extreme case for which both periods are reliably resolved (PAR~$= -3$) and the harmonic structure permits a confident mode classification. With $P_1 = 71$~s, $H = 28.5$, and an estimated diameter of $\sim 7$~m for an albedo $p_V = 0.15$, 2025~FB8 is, to our knowledge, the fastest and smallest tumbling asteroid ever characterised at this level. Previously identified tumblers with two reliably resolved periods have $P_1$ of at least a few minutes and diameters of tens to hundreds of metres. 2025~FB8 extends this boundary into the sub-10~m regime, where the structural integrity and excitation history of these bodies remain essentially unexplored.

Among the 14 PAR~$= -3$ objects, 8 satisfy additional harmonic-quality criteria that allow a tentative dynamical mode classification (Sect.~\ref{sec:lamsam}). All 8 are consistent with LAM rotation, in which the body tumbles with its long axis precessing around the angular momentum vector. No SAM candidate meets the full set of reliability criteria, although one PAR~$= -2$ object (2024~SF2) shows a harmonic signature tentatively consistent with SAM. With only 8 classified objects drawn from a single viewing geometry per target, no quantitative inference about the true LAM/SAM ratio in this population is warranted. The small sample, the single viewing geometry per target and the unknown relative detection efficiency for LAM and SAM prevent this result from being interpreted as a measurement of the intrinsic mode ratio.

\begin{figure*}[t!]
\centering
\begin{minipage}{0.49\textwidth}
    \centering
    \includegraphics[width=\textwidth]{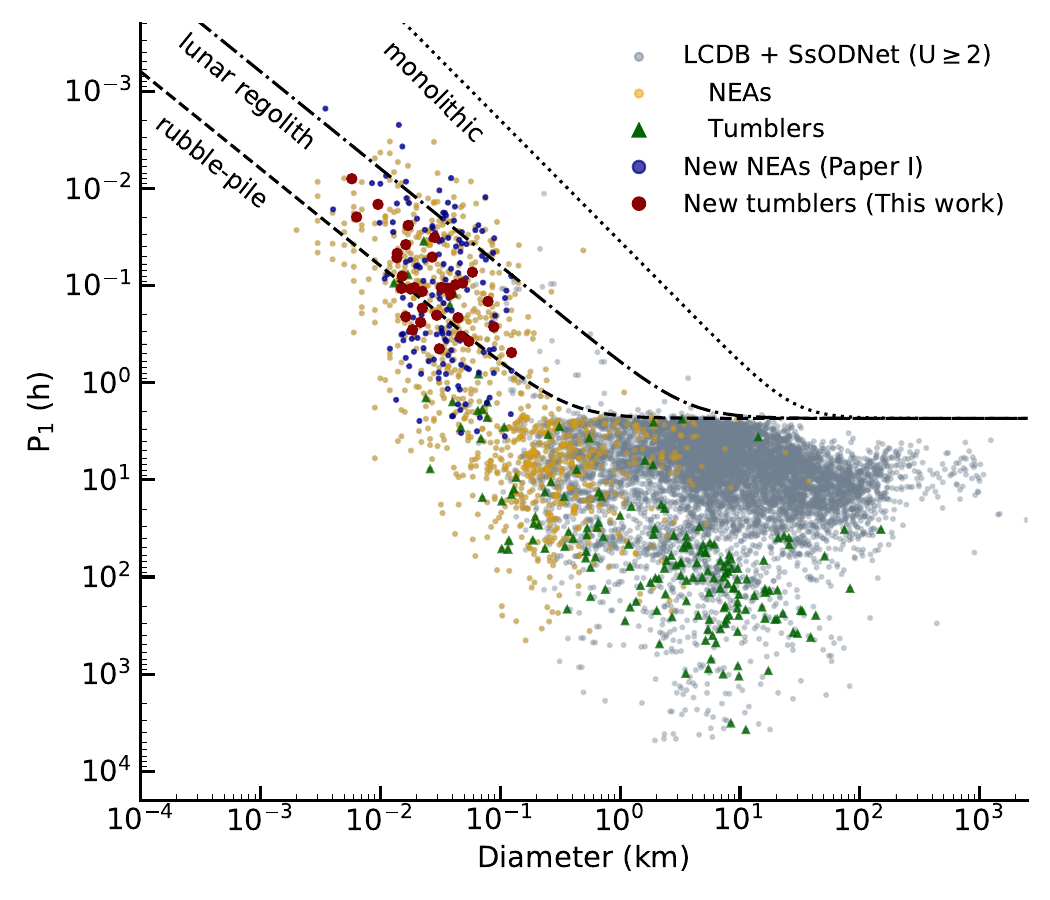}
\end{minipage}
\hfill
\begin{minipage}{0.49\textwidth}
    \centering
    \includegraphics[width=\textwidth]{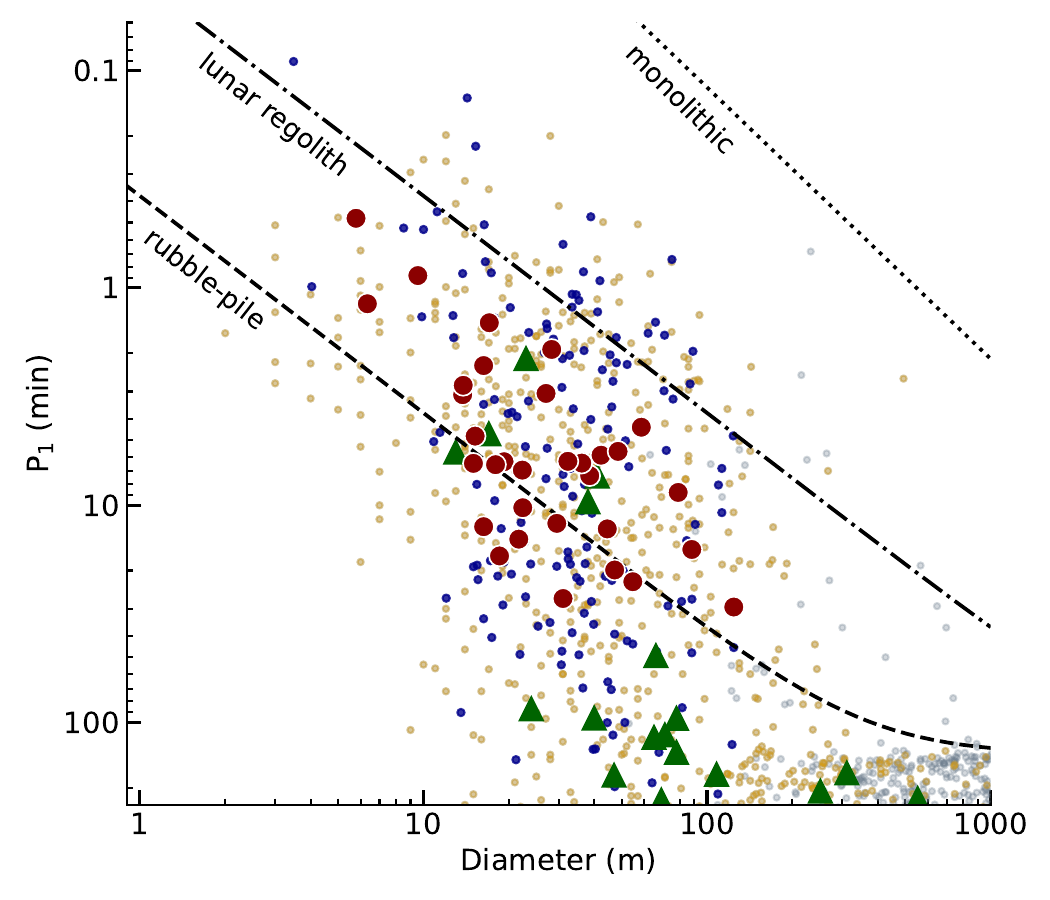}
\end{minipage}
\caption{Period $P_1$ as a function of diameter for the objects in this work and comparison samples. Left: Full distribution. Right: Zoom into the sub-kilometre regime. In both panels, grey points are asteroids from the LCDB and SsODNet with $U \geq 2$; orange points are NEAs from those catalogues; blue circles are the 131 PA fast rotators from \cite{Alarcon2026}; green triangles are tumblers already listed in LCDB or SsODNet; and red circles are the 32 new tumblers presented in this work. For the new tumblers, $P_1$ is the shorter of the two bivariate periods and does not necessarily correspond to the physical rotation period. Diameters are estimated from absolute magnitudes assuming $p_V = 0.15$. Cohesion curves are overplotted: each curve marks the minimum period sustainable by a body of the given internal structure (weak rubble pile, $C \sim 10$~Pa; lunar regolith, $C \sim 1$~kPa; monolithic, $C \propto D^{-0.5}$), so objects with periods shorter than the corresponding limit require at least that level of cohesive strength.}
\label{fig:spin_diagram}
\end{figure*}

\subsection{Period--size distribution}
\label{sec:results_pd}

The structural implications of fast rotation for small NEAs were extensively discussed in \citetalias{Alarcon2026}, where we showed that a substantial fraction of our sample lies above the cohesionless rubble-pile spin limit and therefore requires some degree of internal cohesion. Here we examine how the 32 confirmed tumblers populate the period--size diagram and what this implies for their internal structure.

Figure~\ref{fig:spin_diagram} shows the period $P_1$ as a function of estimated diameter $D$ for the 32 confirmed tumblers (red circles), together with the PA fast rotators from \citetalias{Alarcon2026} (blue), the LCDB and Solar System Open Database Network (SsODNet) NEA sample with $U \geq 2$ (orange), and the tumblers already catalogued in those databases (green triangles). For each tumbler we plot $P_1$---the shorter of the two recovered periods---since it provides the most direct comparison with the single-period rotation rates of the PA sample. We note, however, that $P_1$ is not necessarily the rotation period in the physical sense: without a full spin-state model it cannot be determined whether $P_1$ corresponds to $P_\phi$ or $P_\psi$. The cohesion curves and the diameter convention ($p_V = 0.15$) are the same as in \citetalias{Alarcon2026}: each curve marks the maximum spin rate (minimum period) sustainable by a body with the corresponding internal structure, so objects lying below a given curve require at least that level of cohesive strength.

The tumblers detected in this work occupy a region of the period--size diagram that is almost entirely empty in the existing literature. Previously identified tumblers (green triangles) cluster at long periods and large diameters; the small-NEA fast-rotator regime ($D \lesssim 100$~m, $P_1 \lesssim 1$~h) was essentially unpopulated. Our sample fills this gap with 32 confirmed NPA rotators spanning diameters from $\sim 7$ to $\sim 140$~m and periods from 0.48 to 29.5~min. The zoom panel of Fig.~\ref{fig:spin_diagram} highlights this regime: tumblers and PA fast rotators span the same $P$--$D$ region with no apparent segregation between the two spin states.

Regarding the cohesion limits derived in \citetalias{Alarcon2026}, 23 of the 32 tumblers lie above the weak rubble-pile spin limit and 9 fall below it; all 32 lie below the lunar-regolith limit. These numbers should be interpreted with caution: the position of each object relative to the cohesion curves depends on its estimated diameter, which carries significant uncertainty through the assumed albedo and the absolute magnitude. Additionally, the cohesion curves give the maximum spin rate sustainable by a body of given strength, strictly applicable to the rotation period $P_\psi$. Since $P_1$ can correspond to the precession period ($P_\phi$) rather than the rotation for some objects, the structural interpretation of individual placements in the diagram should be treated accordingly. Bearing these caveats in mind, the tumbler population appears to occupy the same structural domain as the PA fast rotators of \citetalias{Alarcon2026}: a regime where cohesive forces are required but moderate regolith-level strength suffices. Notably, no tumbler exceeds the lunar-regolith spin limit, indicating that none of them requires a monolithic interior to sustain its current spin rate.

\subsection{Prevalence of NPA rotation}
\label{sec:results_prevalence}

To estimate the prevalence of NPA rotation as a function of size, we considered as our parent population the 156 fast rotators ($P < 2.2$~h) identified in \citetalias{Alarcon2026}, together with the 34 tentative fast rotators reported in that survey. To ensure consistency in the observational baseline, we restricted the tumbler sample for this analysis to the 25 objects observed within the same time window (up to 2025 December 12). The remaining seven tumblers were observed in subsequent months and would otherwise bias the comparison. The global tumbler fraction is $f_{\rm tumb} = 16.0\%$ (25 out of 156 confirmed fast rotators, or $13.2\%$ if tentative fast rotators are included in the denominator). Roughly one in six fast rotators in our survey therefore shows statistically significant departures from single-period rotation.

This global fraction suggests a possible dependence on size. Objects with $H \geq 26$ show $f_{\rm tumb} = 22.5\%$ (9 out of 40 fast rotators), compared to $13.8\%$ at $H < 26$ (16 out of 116). Table~\ref{tab:tumbler_prevalence} summarises this breakdown. The split at $H = 26$ coincides with the approximate boundary above which the fast-rotator fraction in \citetalias{Alarcon2026} rises most steeply and where the transition towards strength-dominated interiors becomes most pronounced. A Fisher exact test comparing the tumbler fractions in the two bins gives $p = 0.22$, indicating that the difference is not statistically significant at the current sample size. Including the tentative fast rotators in the denominator gives an essentially identical result. The trend is therefore suggestive but inconclusive and a larger sample would be needed to confirm it.

\begin{table}
\centering
\caption{Prevalence of NPA rotation among fast rotators as a function of absolute magnitude.}
\label{tab:tumbler_prevalence}
\begin{tabular}{cccc}
\hline\hline
Magnitude & $N_{\rm FR}$ & $N_{\rm tumb}$ & $f_{\rm tumb}$ \\
\hline
$H < 26$  & 116 (+26) & 16 & 13.8\% (11.3\%) \\
$H \geq 26$ &  40 (+8)  &  9 & 22.5\% (18.8\%) \\
\hline
Total     & 156 (+34) & 25 & 16.0\% (13.2\%) \\
\hline
\end{tabular}
\tablefoot{$N_{\rm FR}$ is the number of confirmed fast rotators from \citetalias{Alarcon2026} and $N_{\rm tumb}$ the number of confirmed tumblers from this work. Values in parentheses include tentative fast rotators in the denominator.}
\end{table}

An important observational caveat qualifies this result. Our observing runs of 40--120~min are optimised for the detection of fast rotation, which introduces a detection bias against tumblers whose second period ($P_2$) is long relative to the session duration: a tumbler with $P_2 \gtrsim 40$~min may complete fewer than five full cycles within a single run, making its bivariate signature difficult to recover. Since longer $P_2$ values are more common among larger, brighter objects, this bias suppresses $f_{\rm tumb}$ preferentially at the large end of the sample. The same effect explains the relative scarcity of tumblers with $P_1 \gtrsim 40$~min in Fig.~\ref{fig:spin_diagram}. If these periods are more frequent among larger objects, the resulting incompleteness would preferentially reduce the detected fraction in the $H < 26$ bin and could produce or enhance the apparent increase towards smaller sizes. The observed contrast therefore cannot be interpreted as a lower bound on an intrinsic size dependence. Quantifying that dependence requires an assessment of the detection efficiency as a function of period, amplitude, photometric precision and observing baseline.

The apparent increase in $f_{\rm tumb}$ towards smaller sizes is counter-intuitive. Internal dissipation should damp NPA rotation more efficiently in smaller, faster-rotating bodies, so one would expect the tumbler fraction to decrease with $H$. If confirmed, the opposite trend would indicate that the mechanisms that excite and sustain NPA rotation in this size regime operate efficiently enough to replenish the tumbling population against damping. In the following section we examine three pathways that could produce and maintain fast tumblers at these sizes.

\section{Discussion: Pathways to fast-tumbling asteroids}
\label{sec:discussion}
We identify and discuss three possible pathways to the fast-tumbling asteroids that are reported here. These possible pathways are from catastrophic disruption, from YORP disaggregation, or from YORP spin-up dynamics. Each of these is considered in the following, along with possible implications that could be tested in the future. Of course, we also note that there may be other pathways that have not been identified here.

\subsection{Catastrophic disruption} 

It is well known that the lifetime against disruption of asteroids decreases as they decrease in size. This is due to their lower disruption energy values, making them susceptible to smaller impacts having an overall catastrophic result \citep{Holsapple2002}. For the regime of bodies that are considered in this paper, their formation from a catastrophic disruption of a larger body is not out of the question. The catastrophic disruption of a body could shatter it and, if the parent body is relatively modest in size, not provide sufficient gravitational attraction to retain these components. During the disruption process, the individual components would be subjected to shocks and large forces that would cause the components to spin rapidly and immediately enter a tumbling state. This has been studied experimentally as part of a larger series of impact experiments \citep{Giblin1997, Giblin1998}. 

In \citet{Asphaug1998} a number of numerical simulations of the catastrophic disruption of a small asteroid were made, with a specific focus on the asteroid Castalia, using its radar-derived shape model \citep{Hudson1994}. In a follow-on analysis the outcome of a catastrophic disruption of this body was analysed in terms of the shapes, moments of inertia and spin states of the separate shards \citep{Asphaug1999}. It is striking that the rotation states arising from that analysis are in line with the current observations. Fragment sizes from that study ranged from approximately 15~m up to a few larger fragments around 400~m. The spin periods resulting from the shattering ranged from 0.02~h up to a few hours, with the majority of spin periods in the 0.1--1~h range. Finally, the bodies were relatively uniformly distributed between SAM and LAM spin states. These ranges of values are similar to those of the observed bodies. The relative frequency of LAM and SAM states cannot be meaningfully compared. Our mode assignments rest on single-epoch spectra for only eight objects (Sect.~\ref{sec:results_sample}), a sample too small and too weakly constrained to test the near-uniform LAM--SAM partition predicted by the disruption simulations.

Due to the formation of these components, it is likely that they would all be monolithic, given the large stress waves that have created the shattering. Thus, if future observations find that small, fast rotators are in fact rubble piles still (see the next pathway), this would discount the catastrophic disruption pathway for formation. 

\subsection{YORP disaggregation}

A different pathway to a population of tumbling, fast-rotating asteroids arises if we consider the effect of cohesive forces in rubble-pile asteroids. Even at very modest levels of cohesion, which have been documented in earlier asteroid studies \citep{Hirabayashi2014, Hirabayashi2015, Hirabayashi2019}, a rubble-pile asteroid can spin beyond its rotational fission threshold. 
Thus, if a rubble-pile asteroid has a small amount of cohesive strength, due to adhesion between grains or sintering in
the space environment, it is then possible for that rubble pile to spin beyond the usual fission limit for a cohesionless collection of grains \citep{Scheeres2018}. For small bodies the amount of additional strength needed to make a significant increase in the spin rate is quite modest, and has been estimated and measured in previous papers as cited above. The end state of this process can lead to the disaggregation of a rubble pile in a finite period of time if the body spins faster than $\sqrt{2}$ times its usual cohesionless fission limit. For example, spin periods shorter than 1.65~h for a body with density of $2~\mathrm{g\,cm^{-3}}$. Above this spin rate, the components of the body would immediately escape from each other when they fission; however, the components would in general still remain as rubble-pile bodies, just  smaller and with fewer components. 
As the fission process for a cohesive rubble pile is to break in the centre, where the stresses are the highest, it is reasonable that a cohesive rubble pile would break into two, smaller cohesive rubble piles \citep{Hirabayashi2019a}. 
The new rubble-pile bodies would be effectively stronger (since the centripetal acceleration induced stresses will be reduced) and would need to spin faster to fission. 
These bodies could then be spun up again by YORP until they fission, the process repeating until just the monolithic components of the body are left. It should be noted that, even if the progenitor rubble pile is uniformly rotating, the fissioned body will in general immediately enter a tumbling state once it fissions, and thus could also be the source of the observed fast tumblers. The details of this disaggregation process are described and analysed in \citet{Scheeres2018a}. 

The tumblers reported here are consistent with this picture: 23 of the 32 lie above the spin limit of a cohesionless rubble pile, yet none exceeds the lunar-regolith limit (Sect.~\ref{sec:results_pd}). Their spin states therefore require cohesion, but only at the modest levels invoked above, and none of them demands a monolithic interior.
In this model, some fraction of the currently observed fast tumblers would likely be rubble-pile bodies, being accelerated to their next fission.  Testing this model would ideally require showing that one member of this population is a rubble pile, either by observing it fission or through high-resolution observations of the body, for example from a spacecraft. 

\subsection{YORP dynamics on small bodies}
A third pathway arises from the complex rotational dynamics that YORP can induce in small bodies. The traditional picture of YORP spin-up, applicable to bodies of hundreds of metres and larger, envisions a uniformly rotating asteroid whose spin rate increases secularly. At smaller sizes this dynamic becomes considerably richer. This regime has been explored by \citet{Benson2026} for asteroids of 1--10~m, using a formalism originally developed to model the rotational dynamics of defunct Earth-orbiting satellites \citep{Benson2021, Benson2021a, Benson2022}, whose spin states change rapidly under solar radiation pressure. Small asteroids are subject to analogous effects, albeit on longer timescales.

At lower spin rates, the rotational angular momentum of the body can track and precess around the Sun direction, resulting in repeated cycling between uniform and NPA rotation. These slow tumbling cycles can be interrupted if the
period ratio becomes resonant (i.e. $P_{\psi}/P_{\phi}$ approaching a ratio of integers), after which the spin rate increases secularly as the two periods change in lock step to remain resonant. This `run-away resonance' behaviour
has been seen in simulations of Earth-orbiting satellites \citep{Benson2018, Benson2022}. The resonant state does not appear to be stable on its own, so the object would eventually leave the resonance while retaining its high spin
rate and tumbling state until relaxation.

The same simulations reveal two further routes to fast tumbling. In some cases the evolution approaches a limit cycle in which spin-up of the tumbling body continues without the need for a resonance; in others, the body is spun up about its axis of minimum moment of inertia, which reaches faster spin rates for a given torque. In both, once the spin rate becomes high enough, energy dissipation begins to relax the body, ensuring that it remains in a tumbling state throughout the dissipation timescale.

A testable signature of the resonant route is that the dominant light-curve frequencies should all be multiples of a single low frequency, even when the light curve itself is complex. Such a test is beyond the reach of our data. The formal uncertainties on $P_2$ are large enough that the $1\sigma$ interval of a typical period ratio encompasses one or more low-order resonances, so formal consistency with a resonant state carries no diagnostic weight. Discriminating between resonant and non-resonant configurations would require period determinations of considerably higher precision than the single-session baselines available here. Quantifying how often small asteroids become trapped in these evolving spin states will therefore require further analysis, but the documentation of the same behaviours in defunct Earth-orbiting satellites \citep{Benson2021} suggests that they could plausibly account for the states observed here.

\section{Conclusions}\label{sec:conclusions}

We have carried out a systematic search for NPA rotation among the fast-rotating NEAs of \citetalias{Alarcon2026}, using a bivariate Fourier analysis of high-cadence photometry. Our main results are as follows.
\begin{enumerate}
    \item Of the 42 NPA candidates analysed, 32 are confirmed as tumblers: 14 with both periods reliably resolved (PAR~$= -3$) and 18 in which NPA rotation is detected but the second period remains poorly constrained (PAR~$= -2$). This is the largest homogeneous sample of fast tumblers assembled to date.
    \item The recovered periods span $0.48 \leq P_1 \leq 29.5$~min and $2.15 \leq P_2 \leq 46.4$~min, over diameters of $\sim 7$--$140$~m. 2025~FB8, with $P_1 = 71$~s and $D \sim 7$~m, is to our knowledge the fastest and smallest tumbling asteroid for which two periods have been resolved, extending the known tumbler population into the sub-10~m regime.
    \item Eight PAR~$= -3$ objects meet the harmonic quality criteria for a tentative mode assignment, and all eight are consistent with LAM rotation. No object meets the full set of criteria for a SAM solution. With a single viewing geometry per target, this cannot be turned into a constraint on the true LAM/SAM ratio.
    \item The tumbler fraction among fast rotators is $f_{\rm tumb} = 16.0\%$, rising from $13.8\%$ at $H < 26$ to $22.5\%$ at $H \geq 26$. Both figures are lower bounds: our sensitivity to a second period degrades as $P_2$ grows relative to the session duration, and since long $P_2$ values are more common among larger objects, this suppresses $f_{\rm tumb}$ preferentially in the $H < 26$ bin and could account for part or all of the apparent contrast.
    \item In the period--size plane the tumblers occupy the same region as the PA fast rotators, with no apparent segregation between the two spin states. Twenty-three of the 32 lie above the spin limit of a cohesionless rubble pile, and none exceeds the lunar-regolith limit: cohesion is required, but only at modest levels, and no object demands a monolithic interior.
\end{enumerate}

That NPA rotation remains common at the smallest sizes, where internal dissipation should damp it most efficiently, implies that excitation and sustaining mechanisms operate at least as efficiently. We have discussed three pathways consistent with the observations---catastrophic disruption of a larger parent body, disaggregation of a cohesive rubble pile driven by YORP and the complex spin dynamics that YORP induces in metre-scale bodies---which make different predictions for the internal structure of the resulting objects. Distinguishing between them will require either resolved observations of individual fast tumblers or period determinations precise enough to test for spin resonances, neither of which is within reach of single-session survey photometry.

\section*{Data availability}

The derived rotational parameters and the calibrated photometric measurements for all objects presented in this work will be made available to the community through the LCDB. In addition, the survey photometry, including the data analysed in \citetalias{Alarcon2026} and in the present paper, will be made publicly available through AsteroiDB \citep{Alarcon2024}. These tools will allow readers to inspect the light curves, period searches, and fitted solutions for individual objects, including those classified as PAR~$=-2$ and PAR~$=-3$, explore alternative frequency combinations and assess the limitations of the reported solutions.

\begin{acknowledgements}
This article is based on observations made in the Two-meter Twin Telescope (TTT) sited at the Teide Observatory of the Instituto de Astrofísica de Canarias (IAC), that Light Bridges operates in Tenerife, Canary Islands (Spain). The observation time rights (DTO) used for this research were consumed in the PEI "FASTROTA26". This research used storage and computing capacity in ASTRO POC's EDGE computing centre at Tenerife under the form of Indefeasible Computer Rights (ICR), consumed in the PEI "FASTROTA26". Dr Antonio Maudes’s insights in economics and law were instrumental in shaping the development of this work.
M.R.A. and J.L. acknowledge financial support through grant PID2024-160618NB-C22, “Hydrated Minerals and Organic Compounds in Primitive Asteroids 2”, funded by MICIU/AEI/10.13039/501100011033 and by ERDF, EU, and through the Severo Ochoa Centre of Excellence accreditation awarded to the Instituto de Astrof\'{\i}sica de Canarias, grant CEX2025-001609-S, funded by MICIU/AEI/10.13039/501100011033. 
\end{acknowledgements}

\bibliographystyle{aa}
\bibliography{npa}

\end{document}